\documentclass{article}
\usepackage[utf8]{inputenc}

\usepackage{authblk}
\usepackage{orcidlink}

\usepackage{hyperref}

\newcommand*{\email}[1]{\tt\href{mailto:#1}{#1}}

\title{Conversion of Boolean and Integer FlatZinc Builtins to Quadratic or Linear Integer Problems\footnote{The presented work was performed in the EniQmA project funded by the
Federal Ministry for Economic Affairs and Climate Action under the FKZ 01MQ22007A.}}

\author[1]{Armin Wolf\,\orcidlink{0000-0003-3940-0792}}

\affil[1]{Fraunhofer FOKUS, Kaiserin-Augusta-Allee 31, D-10589 Berlin, Germany,
\email{armin.wolf@fokus.fraunhofer.de}}

\date{Version of \today}

\usepackage{amsfonts} 

\usepackage[colorinlistoftodos]{todonotes} 

\newcommand{\f}[1]{\mbox{\textsf{#1}}}

\begin{document}

\maketitle


\begin{abstract}
Constraint satisfaction or optimisation models --- even if they are formulated 
in high-level modelling languages --- need to be reduced into an equivalent format 
before they can be solved by the use of Quantum Computing. In this paper we show how
Boolean and integer FlatZinc builtins over finite-domain integer variables
can be equivalently reformulated as linear equations, linear inequalities or binary
products of those variables, i.e. as finite-domain quadratic integer programs. Those 
quadratic integer programs can be further transformed into equivalent Quadratic 
Unconstrained  Binary Optimisation problem models, i.e. a general format for 
optimisation problems to be solved on Quantum Computers especially on Quantum Annealers.  
\end{abstract}

\section{Introduction}

Our work in the EniQmA project\footnote{see 
\href{https://www.eniqma-quantum.de/}{https://www.eniqma-quantum.de/}} aims to 
transform MiniZinc programs\footnote{see 
\href{https://www.minizinc.org/}{https://www.minizinc.org/}}
defining constraint satisfaction or optimisation problems (CSP/COP) into 
equivalent Quadratic Unconstrained Binary Optimisation (QUBO) problems.
The intention is to offer thereby a tool to support the solving of a large set
of satisfaction and optimisation problems by the use of Quantum Computing. 
In particular, to use Quantum Annealers beyond other solvers in the MiniZinc 
context.  

Thanks to the developers of MiniZinc~\cite{nethercoteMiniZincStandardCP2007} and the
accorcding software tools we can focus on FlatZinc 
programs\footnote{The language FlatZinc is a proper subset of the language MiniZinc.} 
because any MiniZinc program can be transformed automatically by the MiniZinc compiler 
into an equivalent FlatZinc program. A large set of FlatZinc programs are representing 
\emph{quadratic or linear constraint optimisation problems} over finite domain integer variables. 
Therefore we focus in this document on the transformation of Boolean and integer
Flat\-Zinc builtins to finite-domain quadratic integer problems, which can be further
transformed into QUBO problems~\cite{wolf2023automatic}.

\section{Finite-Domain Quadratic Integer Problems} \label{sec:QIP}

Here we focus on \emph{Finite Domain Quadratic Integer Problems 
(QIP(FD))}, i.e. on finite integer optimisation problems of the form
\begin{eqnarray}
    \mbox{minimize} && \sum_{i=1}^n g_i \cdot x_i \label{eqn:orgObj}
\end{eqnarray}
subject to 
\begin{eqnarray}
    \sum_{i=1}^n a_{j,i} \cdot x_i + \sum_{k=1}^m b_{j,k} \cdot y_k + c_j \le 0 
    & & \mbox{for $j = 1, \ldots, p$} \label{eqn:orgIneqs} \\
    \sum_{i=1}^n d_{j,i} \cdot x_i + \sum_{k=1}^m e_{j,k} \cdot y_k + f_j = 0 
    & & \mbox{for $j = p+1, \ldots, q$} \\
    y_k = z_v \cdot z_w  
    & & \mbox{$z_v, z_w \in \{x_1, \ldots, x_n, y_1, \ldots, y_{k-1}\}$} \nonumber \\
    & & \mbox{for $k = 1 \ldots, m$} \label{eqn:orgProd} \\[1ex]
    x_i \in D(x_i) \subset \mathbb{Z} & & \mbox{for $i = 1, \ldots, n$} \\
    y_k \in D(y_k) \subset \mathbb{Z} & & \mbox{for $k = 1, \ldots, m$} \enspace,
    \label{eqn:orgLast}
\end{eqnarray}
where $g_i, a_{j,i}, b_{j,k}, c_j,  d_{j,i}, e_{j,k}, f_j$ are rational or integer values, 
$x_i, y_k$ are uniquely defined variables with non-empty finite domains $D(x_i)$ resp. 
$D(y_k)$. We assume that the finite domains of the variables are \emph{bounds-consistent
integer intervals}, i.e. $D(x_i) = [\min(x_i),\max(x_i)]$ resp. $D(y_k) = 
[\min(y_k),\max(y_k)]$, e.g. according to the pruning rules defined ´
in~\cite{schulteWhenBoundsDomain2001} (Figure~1).

In the special case that the objective in Equation~(\ref{eqn:orgObj}) is missing (or zero) 
the problem is a satisfaction problem and if the binary variable products, i.e. the 
Equations~(\ref{eqn:orgProd}) are missing, the problem is linear.

\section{Representing FlatZinc Builtins}\label{sec:fz}

Here we show how to transform Bool(ean) and integer FlatZinc 
builtins~\cite{FlatZincBuiltinsMiniZinc}, i.e. these FlatZinc constraints into 
equivalent conjunctions  of linear equations and inequalities as well as 
binary variable products, i.e. into QIP(FD) constraints. Therefore, we assume 
that each variable $x$ has an integer domain~$D(x) =
[\min(D(x)), \max(D(x))] = [\min(x), \max(x)]$  where $\min(x), \max(x) 
\in \mathbb{N}$ with $\min(x) \le \max(x)$, either defined by a 
statement {\tt var $\min(x)$..$\max(x)$:~x;} or by {\tt var bool:~x;} in FlatZinc.

Boolean values are represented by~\f{false}  and~\f{true} in FlatZinc, which can be 
explicitly converted to integer values~$0$ and~$1$ by the use of the {\tt bool2int}
bultin. Therefore, we assume that this builtin is (implicitly) applied to each Boolean
variable resulting in a binary variable, i.e. any Boolean value~$b \in \{\f{false}, 
\f{true}\}$ is considered as a binary (integer) value~$b \in \{0, 1\} = [0,1]$.

It should be noted that we do not consider special cases, when we transform
FlatZinc builtins into QIP(FD) constraints, e.g. the fact that $(a = b) \leftrightarrow r$ 
can be reduced to $a=b$ if $r=1$ applies, is ignored in the following considerations. 

\subsection{Representing Integer FlatZinc Builtins}

\subsubsection{The predicate {\tt array\_int\_element}}
\label{sec:fz:array_int_element}
The semantics of {\tt array\_int\_element(var int:~i, array~[int]~of~int:~a, 
var~int:~c)} is $c = a_i$, i.e. the value of the integer variable $c$ is the 
$i$-th value in the integer array~$a$ where $i$ is an integer index variable,
$a = a_1, \ldots, a_n$ are integer values.  Then, we can restrict the domains of 
the variables: 
\begin{eqnarray*}
    i & \in &  [\max(1,\min(i)), \min(n,\max(i))] \\ 
    c & \in & [\min_{i \in [\min(i), \max(i)]} \min(a_i), 
        \max_{i \in[\min(i), \max(i)]} \max(a_i)] 
\end{eqnarray*}

and the predicate can be represented by linear equations:
\begin{eqnarray}
    1 = \sum_{j=\min(i)}^{\max(i)} b_j^i \label{eq:int:one-hot_1}\:,
    \qquad
    i = \sum_{j=\min(i)}^{\max(i)} j \cdot b_j^i 
    \qquad
    c = \sum_{j=\min(i)}^{\max(i)} a_j \cdot b_j^i\:,  \label{eq:int:one-hot}
\end{eqnarray}
where $b_{\min(i)}^i \in [0,1] , \ldots, b_{\max(i)}^i \in [0,1]$ are new auxiliary 
binary variables.

It should be noted that the first two equations in~(\ref{eq:int:one-hot}) define 
a \emph{one-hot-encoding} of the integer variable~$i$. This should be respected
when transforming the resulting QIP(FD) into a QUBO problem. 

\subsubsection{The predicate {\tt array\_int\_maximum}}
\label{sec:fz:array_int_maximum}
The semantics of {\tt array\_int\_maximum(var~int:~m, array~[int]~of~var~int: x)} 
is $m = \max_i x_i$, i.e. the value of the integer variable $m$ is 
the maximum value of the values of the integer variables array~$x_1, \ldots, x_n$.
Then we can restrict the domains of the variables:
\begin{eqnarray*}
  m & \in & [\max(\min(m), \max_{i=1,\ldots,n}\min(x_i)), \min(\max(m),
    \max_{i=1,\ldots,n}\max(x_i))] \\
  x_i & \in & [\min(x_i), \min(\max(x_i),\max(m))] \quad\mbox{for $i=1,\ldots,n$.}
\end{eqnarray*}  
This can be presented by linear equations and inequalities as well as binary variable
products:
\begin{eqnarray*}
    1 = \sum_{j=1}^{n} b_j & \mbox{and} &  m = \sum_{j=1}^{n} z_j \\
    z_j = x_j \cdot b_j & \mbox{and} &  m  \ge x_j \quad\mbox{for $j=1, \ldots, n$.}
\end{eqnarray*}
where $b_{1} \in [0,1] , \ldots, b_{n} \in [0,1]$ are new auxiliary binary
variables and $z_1 \in [\min(0, \min(a_1)), \max(a_1)], \ldots, z_n \in [\min(0, 
\min(a_n)), \max(a_n)]$ are new auxiliary integer variables.

\subsubsection{The predicate {\tt array\_int\_minimum}}
\label{sec:fz:array_int_maximum}
The semantics of {\tt array\_int\_minimum(var int:~m, array [int] of 
var int: x)} is $m = \min_i x_i$, i.e. the value of the integer variable $m$ is 
minimum value of the integer variables array~$x_1, \ldots, x_n$.
Then we can restrict the domains of the variables:
\begin{eqnarray*}
  m & \in & [\max(\min(m), \min_{i=1,\ldots,n}\min(x_i), \min(\max(m),
    \min_{i=1,\ldots,n}\max(x_i)] \\
  x_i & \in & [\max(\min(x_i),\min(m)), \max(x_i)] \quad\mbox{for $i=1,\ldots,n$.}
\end{eqnarray*}  
This can be presented by linear equations and inequalities as well as binary variable
products:
\begin{eqnarray*}
    1 = \sum_{j=1}^{n} b_j & \mbox{and} & m = \sum_{j=1}^{n} z_j \\
    z_j = x_j \cdot b_j &\mbox{and} & m \le x_j \quad\mbox{for $j=1, \ldots, n$,}
\end{eqnarray*}
where $b_{1} \in [0,1] , \ldots, b_{n} \in [0,1]$ are new auxiliary binary
variables and and $z_1 \in [\min(0, \min(a_1)), \max(a_1)], \ldots, z_n \in [\min(0, 
\min(a_n)), \max(a_n)]$ are new auxiliary integer variables.

\subsubsection{The predicate {\tt array\_var\_int\_element}}

The semantics of {\tt array\_var\_int\_element(var\:int:\:i, array\:[int]\:of\:var:\:a, 
var\:int:\:c)} is $c = a_i$, i.e. the value of the integer variable $c$ is 
equal to the value of $i$-th variable in the integer variable array~$a$ where $i$ 
is an integer index variable, $a = a_1, \ldots, a_n$ are integer variables. We can
restrict the domain of~$i$ to $i \in [\max(1,\min(i)), \min(n,\max(i))]$ and the domain of~$c$ to $c \in [\min_{i \in [\min(i), \max(i)]} \min(a_i), 
\max_{i \in[\min(i), \max(i)]} \max(a_i)]$. Then the predicate can be represented 
by linear equations and binary variable products:
\begin{eqnarray*}
    && 1 = \sum_{j=\min(i)}^{\max(i)} \beta_j^i \label{eq:int:one-hot_1}\:,
    \qquad
    i = \sum_{j=\min(i)}^{\max(i)} j \cdot \beta_j^i 
    \qquad
    c = \sum_{j=\min(i)}^{\max(i)} z_j\:, \label{eq:var_int:one-hot} \\
    && z_j = a_j \cdot \beta_j^i \quad\mbox{for $j=\min(i), \ldots,  \max(i)$,} 
\end{eqnarray*}
where $\min(i)$ and $\max(i)$ are the updated boundaries of the restricted domain of~$i$ 
(see above), $\beta_{\min(i)}^i \in [0,1] , \ldots, \beta_{\max(i)}^i \in [0,1]$ are new 
auxiliary binary variables, and the variables $z_{j} \in [\min(0, \min(a_{j})), \max(0,\max(a_{j}))]$ with $j=\min(i), \ldots, \max(i)$ are new auxiliary integer variables.

It should be noted that the first two equations in~(\ref{eq:var_int:one-hot}) 
define a \emph{one-hot encoding} of the integer variable~$i$. This should be respected
when transforming the resulting QIP(FD) into a QUBO problem. 


\subsubsection{The predicate {\tt int\_abs}}
\label{sec:fz:int_abs}
The semantics of {\tt int\_abs(var int:~x, var int:~y)} is 
$y = |x|$, i.e. the value of the integer variable~$y$ is equal to the absolute 
value of the integer variable~$x$. We can restrict the domain of~$y$ to $y \in [\max(0,\min(y)), \max(0,\max(y))]$ reflecting $y \ge 0$. Then the predicate can be represented by a linear equation and a binary variable product:
\begin{eqnarray*}
    y = x - 2\cdot z & \mbox{and} & z = b \cdot x 
\end{eqnarray*}
where 
$z \in [\min(0, \min(x)), \max(0,\max(x))]$ is a new auxiliary integer variable 
and $b \in [0,1]$ is a new auxiliary binary variable.

\subsubsection{The predicate {\tt int\_div}}
\label{sec:fz:int_div}
The semantics of {\tt int\_div(var int:~n, var int:~d, var int:~q)} is as 
follows: The quotient~$q$ produced for operand values~$n$ and~$d$ is an integer 
value~$q$ whose magnitude is as large as possible while satisfying $|d \cdot q| 
\le |n|$. Moreover, $q$ is positive when $|n| \ge |d|$ and $n$ and $d$ have 
the same sign, but $q$ is negative when $|n| \ge |d|$ and $n$ and $d$ have 
opposite signs. Obviously, the constraint is not satisfiable for $d=0$ 
(division by zero) such that $d \ne 0$ respective $|d| >0$ must hold
(see below) and it holds that $q=0$ when $|n| < |d|$, in particular if $n = 0$.
In all other cases $|n| - |d| < |d \cdot q| \le |n|$ must hold such that we
distinguish the following four cases:
\begin{enumerate}
    \item $n > 0$ and $d > 0$: It must hold that 
        $n - d < d \cdot q$ and $d \cdot q \le n$.
    \item $n < 0$ and $d < 0$: It must hold that 
        $n - d > d \cdot q$ and $d \cdot q \ge n$.
    \item $n > 0$ and $d < 0$: It must hold that 
        $n + d < d \cdot q$ and $d \cdot q \le n$.
    \item $n < 0$ and $d > 0$: It must hold that 
        $n + d > d \cdot q$ and $d \cdot q \ge n$.
\end{enumerate}
In the general case when neither $n \ge 0$, $n \le 0$, $d > 0$ nor $d < 0$ holds 
in advance, we introduce two additional binary variables~$\alpha \in [0,1]$ 
and~$\beta \in [0,1]$ to encode the preconditions of the four cases:
\begin{itemize}
    \item Let~$\alpha=1$ if $n > 0$ and $\alpha=0$ if $n < 0$:
        \begin{eqnarray*}
            n & > & (1 -\alpha) \cdot (\min(n)-1) \\
            n & < & \alpha \cdot (\max(n)+1)
        \end{eqnarray*}
    \item Let~$\beta=1$ if $d > 0$ and $\beta=0$ if $d < 0$:
        \begin{eqnarray*}
            d & > & (1 -\beta) \cdot (\min(d)-1) \\
            d & < & \beta \cdot (\max(d)+1)
        \end{eqnarray*}
\end{itemize}
Using this encoding we are able to encode the four cases in one conjunction
of linear or quadratic inequalities
\begin{eqnarray*}
    n > (1 -\alpha) \cdot (\min(n)-1)
    & \land &  n < \alpha \cdot (\max(n)+1) \\
    d > (1 -\beta) \cdot (\min(d)-1)
    & \land & d < \beta \cdot (\max(d)+1) \\
    d \cdot q \le n + (1 - \alpha) \cdot M 
    & \land & d \cdot q \ge n - \alpha \cdot M \\ 
    n - d < d \cdot q + (1 -\alpha\cdot\beta) \cdot M 
    & \land & n - d > d \cdot q - (1 - (1-\alpha)\cdot(1-\beta)) \cdot M  \\
    n + d < d \cdot q + (1 -\alpha\cdot(1-\beta)) \cdot M 
    & \land & n + d > d \cdot q - (1 - \beta\cdot(1-\alpha)) \cdot M 
\end{eqnarray*}
where $M$ is a sufficiently large integer value. These inequalities 
can be further simplified by introducing an auxiliary integer variable~$p$ 
representing the product $d \cdot q$ where 
\begin{eqnarray*}
    D(p) & = & [\min(\min(d)\cdot\min(q),\min(d)\cdot\max(q), 
        \max(d)\cdot\min(q), \max(d)\cdot\max(q)), \\
    & & \:\max(\min(d)\cdot\min(q),\min(d)\cdot\max(q),
        \max(d)\cdot\min(q), \max(d)\cdot\max(q))]
\end{eqnarray*}
and an auxiliary binary variable~$\gamma$ representing the 
product~$\alpha\cdot\beta$:
\begin{eqnarray*}
    n \ge \min(n) -\alpha \cdot (\min(n)-1)
    & \land &  n \le \alpha \cdot (\max(n)+1) -1 \\
    d \ge \min(d) -\beta \cdot (\min(d)-1)
    & \land & d \le \beta \cdot (\max(d)+1) -1 \\
    p \le n + M - \alpha \cdot M 
    & \land & p \ge n - \alpha \cdot M \\ 
    n - d \le p + M - \gamma \cdot M -1
    & \land & n - d \ge p - \alpha \cdot M - \beta \cdot M + \gamma \cdot M +1 \\  
    n + d \le p + M - \alpha \cdot M + \gamma \cdot M -1
    & \land & n + d \ge p  - M + \beta \cdot M + \gamma \cdot M +1 \\ 
    p = d \cdot q & \land & \gamma = \alpha \cdot \beta  
\end{eqnarray*}
The integer value~$M$ is sufficiently large if 
\begin{eqnarray*}
    \max(n-p) & \le & M \\
    \min(p-n) & \ge & -M \\
    \max(n-d-p) & \le & M-1 \\
    \min(n-d-p) & \ge & 1-M \\
    \max(n+d-p) & \le & M-1 \\
    \min(n+d-p) & \ge & 1-M 
\end{eqnarray*}
respective
\begin{eqnarray*}
    \max(n) - \min(p) & \le & M \\
    - \min(p) + \max(n) & \le & M \\
    \max(n) - \min(d) -\min(p) + 1 & \le & M \\
    -\min(n) + \max(d) + \max(p) + 1 & \le & M \\
    \max(n) + \max(d) - \min(p) + 1 & \le & M \\
    - \min(n) - \min(d) + \max(p) +1 & \le & M 
\end{eqnarray*}
i.e. $M$ must be at least as large as the maximum value of the right-hand-sides 
of these inequalities.

\subsubsection{The Predicate {\tt int\_eq}}
The semantics of {\tt int\_eq(var int:~a, var int:~b)} is $a = b$, i.e. the value
of the integer variable~$a$ is equal to the value of the integer variable~$b$
which is a linear equation.

\subsubsection{The predicate {\tt int\_eq\_reif}}
\label{sec:fz:int_eq_reif}
The semantics of {\tt int\_eq\_reif(var int:~a, var int:~b, var bool:~r)}
is $a = b \leftrightarrow r$, i.e. the value of the Boolean (binary) variable $r$ 
is 1 if the values integer variables~$a$ and~$b$ are equal and 0 if they are
different. 
This can be represented by binary variable products, linear equations and 
linear inequalities:
\begin{eqnarray*}
   x & = & r \cdot a\\
   y & = & r \cdot b \\
   x & = & y \\
   (1-r) & \le & (1-r)\cdot(|a| + |b|)
\end{eqnarray*}
where $x \in [\min(0, \min(a)), \max(0,\max(a))]$ and $y \in [\min(0, \min(b)), 
\max(0,\max(b))]$ are new auxiliary integer variables. 
The last inequality is necessary, to cover the case where 
$a = b = 0$, i.e. to force that $r=1$. The modelling of the absolutes $|a|$ 
and $|b|$ is already described in Section~\ref{sec:fz:int_abs}, such that the resulting constraints are:
\begin{eqnarray*}
    x & = & r \cdot a\\
    y & = & r \cdot b \\
    x & = & y \\
    r & \ge & s + t - u - v + 1\\
    s & = & r \cdot u \\
    t & = & r \cdot v \\
    u & = & a - 2\cdot p \\
    p & = & c \cdot a \\
    v & = & b - 2\cdot q \\
    q & = & d \cdot b
\end{eqnarray*}
where $c\in [0,1]$ and $d \in [0,1]$ are new auxiliary binary variables and where
\begin{eqnarray*}
    \alpha(u) & = & \left\{\begin{array}{rl}
        \min(a) & \mbox{if $0 \le \min(a)$,} \\
        -\max(a) & \mbox{if $\max(a) \le 0$,} \\
        0 & \mbox{otherwise.}
    \end{array}\right. \\
    \beta(u) & = & \left\{\begin{array}{rl}
        \max(a) & \mbox{if $0 \le \min(a)$,} \\
        -\min(a) & \mbox{if $\max(a) \le 0$,} \\
        \max(-\min(a),\max(a)) & \mbox{otherwise.}
    \end{array}\right. \\
    \alpha(v) & = & \left\{\begin{array}{rl}
        \min(b) & \mbox{if $0 \le \min(b)$,} \\
        -\max(b) & \mbox{if $\max(b) \le 0$,} \\
        0 & \mbox{otherwise.}
    \end{array}\right. \\
    \beta(v) & = & \left\{\begin{array}{rl}
        \max(b) & \mbox{if $0 \le \min(b)$,} \\
        -\min(b) & \mbox{if $\max(b) \le 0$,} \\
        \max(-\min(b),\max(b)) & \mbox{otherwise.}
    \end{array}\right.
\end{eqnarray*}
are integer values and the variables $u \in [\alpha(u), \beta(u)]$, 
$v \in [\alpha(v), \beta(v)]$ as well as
$s \in [\min(0, \min(u)), \max(0,\max(u))]$, 
$t \in [\min(0, \min(v)), \max(0,\max(v))]$,
$p \in [\min(0, \min(a)), \max(0,\max(a))]$
and $q \in [\min(0, \min(b)), \max(0,\max(b))]$
are new auxiliary integer variables.

\subsubsection{The Predicate {\tt int\_le}}
The semantics of {\tt int\_le(var int:~a, var int:~b)} is $a \le b$, i.e. the value
of the integer variable~$a$ is less than or equal to the value of the integer 
variable~$b$ which is a linear inequality.

\subsubsection{The predicate {\tt int\_le\_reif}}
\label{sec:fz:int_le_reif}
The semantics of {\tt int\_le\_reif(var int:~a, var int:~b, var bool:~r)}
is $a \le b \leftrightarrow r$, i.e. the value of the Boolean (binary) variable $r$ 
is 1 if the value of the integer variable~$a$ is less than or equal to the value
of the integer variable~$b$ and 0 if the value of the integer variable~$a$ is 
greater than the value of the integer variable~$b$
This can be represented by linear inequalities:
\begin{eqnarray*}
    a & \le & b + (1-r) \cdot (\max(a) - \min(b)) \\
    b & \le & a - 1 + r \cdot (\max(b) - \min(a) + 1) \enspace.
\end{eqnarray*}

\subsubsection{The predicate {\tt int\_lin\_eq}}
The semantics of {\tt int\_lin\_eq(array [int] of int:~as, array [int] of var
int:~bs, int:~c)} is $\sum_i as_i \cdot bs_i = c$, i.e. the sum of the values of 
the integer variables~$bs_1, \ldots, bs_n$ weighted by the integer values $as_1, \ldots, 
as_n$ --- assuming that both arrays have the same length~$n$ --- is equal to the 
integer value~$c$. This is a linear equation.

\subsubsection{The predicate {\tt int\_lin\_eq\_reif}}
\label{sec:fz:int_lin_eq_reif}
The semantics of {\tt int\_lin\_eq\_reif(array [int] of int:~as, array [int] of 
var int:~bs, int:~c, var bool: r))} is $\sum_i as_i \cdot bs_i = c 
\leftrightarrow r$, i.e. the value of the Boolean (binary) variable $r$ is~1 if 
the sum of the values of the variables~$bs_1, \ldots, bs_n$ weighted by the integer 
values $as_1, \ldots, as_n$ --- assuming that both arrays have the same length~$n$ 
--- is equal to the integer value~$c$ and the value of~$r$ is~0 in all other cases.
Following the approach in Section~\ref{sec:fz:int_eq_reif} this can be represented
by binary variable products, linear equations and linear inequalities:
\begin{eqnarray*}
    x & = & \sum_{i=1}^n as_i \cdot bs_i - c \\
    |x| & \le & \max(|x|) \cdot (1-r) \\
    1 - r & \le & |x|
\end{eqnarray*}
where $x \in [l(x), u(x)]$ with
\begin{eqnarray*}
    l(x) & = & \sum_{i=1 \land as_{i} < 0}^n as_{i} \cdot\max(bs_i) 
        + \sum_{i=1 \land as_{i} > 0}^n as_{i} \cdot\min(bs_i) - c \\
    u(x) & = & \sum_{i=1 \land as_{i} < 0}^n as_{i} \cdot\min(bs_i) 
        + \sum_{i=1 \land as_{i} > 0}^n as_{i} \cdot\max(bs_i) - c 
\end{eqnarray*}
and where the modelling of the absolute~$|x|$ described 
in Section~\ref{sec:fz:int_abs} is used.

\subsubsection{The predicate {\tt int\_lin\_le}}
The semantics of {\tt int\_lin\_le(array\:[int]\:of int:~as, array\:[int]\:of var
int:~bs, int:~c)} is $\sum_i as_i \cdot bs_i \le c$, i.e. the sum of the values 
of the variables~$bs_1, \ldots, bs_n$ weighted by the integer values $as_1, \ldots,
as_n$ --- assuming that both arrays have the same length~$n$ --- is less than or 
equal to the  integer value~$c$ which is a linear inequality.

\subsubsection{The predicate {\tt int\_lin\_le\_reif}}
The semantics of {\tt int\_lin\_le\_reif(array [int] of int:~as, array [int] of 
var int:~bs, int:~c, var bool: r))} is $\sum_i as_i \cdot bs_i \le c 
\leftrightarrow r$, i.e. the value of the Boolean (binary) variable $r$ is~1 if 
the sum of the values of the variables~$bs_1, \ldots, bs_n$ weighted by the integer
values $as_1, \ldots, as_n$ --- assuming that both arrays have the same length~$n$ 
--- is less than or equal to the  integer value~$c$ and the value of~$r$ is~0 in all
other cases. Following the approach in Section~\ref{sec:fz:int_le_reif} this
can be represented by linear equations and linear inequalities:
\begin{eqnarray*}
    x & = & \sum_{i=1}^n as_i \cdot bs_i - c\\
    x & \le & (1-r) \cdot \max(x) \\
    1 & \le & x + r - r \cdot \min(x)
\end{eqnarray*}
where $x \in [l(x), u(x)]$ is a new auxiliary integer variable with
\begin{eqnarray*}
    l(x) & = & \sum_{i=1 \land as_{i} < 0}^n as_{i} \cdot\max(bs_i) 
        + \sum_{i=1 \land as_{i} > 0}^n as_{i} \cdot\min(bs_i) - c \\
    u(x) & = & \sum_{i=1 \land as_{i} < 0}^n as_{i} \cdot\min(bs_i) 
        + \sum_{i=1 \land as_{i} > 0}^n as_{i} \cdot\max(bs_i) - c \enspace. 
\end{eqnarray*}

\subsubsection{The predicate {\tt int\_lin\_ne}}
The semantics of {\tt int\_lin\_ne(array\:[int]\:of int:~as, array\:[int]\:of var
int:~bs, int:~c)} is $\sum_i as_i \cdot bs_i \ne c$, i.e. the sum of the values of 
the variables~$bs_1, \ldots, bs_n$ weighted by the integer values $as_1, \ldots, 
as_n$ --- assuming that both arrays have the same length~$n$ --- is not equal to the 
integer value~$c$. Following the approach in Section~\ref{sec:fz:int_abs} this can 
be represented by linear equations, linear inequalities and binary variable products:
\begin{eqnarray*}
    x = \sum_{i=1}^n as_i \cdot bs_i - c & \mbox{and} & 1 \le |x|   
\end{eqnarray*}
where $x \in [l(x), u(x)]$ is a new auxiliary integer variable with
\begin{eqnarray*}
    l(x) & = & \sum_{i=1 \land as_{i} < 0}^n as_{i} \cdot\max(bs_i) 
        + \sum_{i=1 \land as_{i} > 0}^n as_{i} \cdot\min(bs_i) - c \\
    u(x) & = & \sum_{i=1 \land as_{i} < 0}^n as_{i} \cdot\min(bs_i) 
        + \sum_{i=1 \land as_{i} > 0}^n as_{i} \cdot\max(bs_i) - c 
\end{eqnarray*}
and $|x| \in [\max(1, \min(|\min(x)|,|\max(x)|)), \max(|\min(x)|, |\max(x)|)]$.
For the representation of $|x|$ see Section~\ref{sec:fz:int_abs}.

\subsubsection{The predicate {\tt int\_lin\_ne\_reif}}\label{sec:fz:int_lin_ne_reif}
The semantics of {\tt int\_lin\_ne\_reif(array [int] of int:~as, array [int] of 
var int:~bs, int:~c, var bool: r))} is $\sum_i as_i \cdot bs_i \ne c 
\leftrightarrow r$, i.e. the value of the Boolean (binary) variable $r$ is~1 if 
the sum of the values of the variables~$bs_1, \ldots, bs_n$ weighted by the integer 
values $as_1, \ldots, as_n$ --- assuming that both arrays have the same length~$n$ 
--- is not equal to the integer value~$c$ and the value of~$r$ is~0 in all other cases. 
Following the approaches in Section~\ref{sec:fz:int_eq_reif} and
Section~\ref{sec:fz:int_lin_eq_reif} this can be represented by 
\begin{eqnarray*}
    x & = & \sum_{i=1}^n as_i \cdot bs_i - c \\
    x & = &  r \cdot x \\
    r & \le & r \cdot |x|
\end{eqnarray*}
respective
\begin{eqnarray*}
    x & = & \sum_{i=1}^n as_i \cdot bs_i - c \\
    x & = &  y \\
    y & = & r \cdot x \\
    r & \le &  z \\
    z & = & r \cdot |x|
\end{eqnarray*}

where $x \in [l(x), u(x)]$ is a new auxiliary integer variable with
\begin{eqnarray*}
    l(x) & = & \sum_{i=1 \land as_{i} < 0}^n as_{i} \cdot\max(bs_i) 
        + \sum_{i=1 \land as_{i} > 0}^n as_{i} \cdot\min(bs_i) - c \\
    u(x) & = & \sum_{i=1 \land as_{i} < 0}^n as_{i} \cdot\min(bs_i) 
        + \sum_{i=1 \land as_{i} > 0}^n as_{i} \cdot\max(bs_i) - c
\end{eqnarray*}
and $|x| \in [l(|x|), u(|x|)]$ with
\begin{eqnarray*}
    l(|x|) & = & \left\{\begin{array}{rl}
                    0 & \mbox{if $l(x) \le 0$} \\
                    l(x) & \mbox{otherwise}
                \end{array}\right. \\
    u(|x|) & = &  \max(|l(x)|, |u(x)|) 
\end{eqnarray*}
$y \in [\min(0, \min(x)), \max(0,\max(x))]$,
$z \in [\min(0, \min(|x|)), \max(0,\max(|x|))]$
are new auxiliary integer variables.
For the representation of $|x|$ see Section~\ref{sec:fz:int_abs}.


\subsubsection{The Predicate {\tt int\_lt}}
The semantics of {\tt int\_lt(var int:~a, var int:~b)} is $a < b$, i.e. the value
of the integer variable~$a$ is less than the value of the integer variable~$b$. 
This can be represented by a linear inequality: 
\begin{eqnarray*}
    a & \le & b - 1 \enspace. 
\end{eqnarray*}

\subsubsection{The predicate {\tt int\_lt\_reif}}
The semantics of {\tt int\_lt\_reif(var int: a, var int: b, var bool: r)}
is $a < b \leftrightarrow r$, i.e. the value of the Boolean (binary) variable $r$ 
is 1 if the value of the integer variable~$a$ is less than the value
of the integer variable~$b$ and 0 if the value of the integer variable~$a$ is 
greater than or equal to the value of the integer variable~$b$.
This can be represented by linear inequalities:
\begin{eqnarray*}
    a & \le & b -1 + (1-r) \cdot (\max(a) - \min(b) + 1) \\
    b & \le & a + r \cdot (\max(b) - \min(a)) \enspace.
\end{eqnarray*}

\subsubsection{The predicate {\tt int\_max}}
The semantics of {\tt int\_max(var int:~a, var int:~b, var int:~c)} is the equation 
$\max(a, b) = c$, i.e. the value of the integer variable~$c$ is the maximum
of the values of the integer variables~$a$ and~$b$. This can be represented by
the linear inequalities:
\begin{eqnarray*}
    a & \le & c \\
    b & \le & c \\
    c & \le & a + r \cdot (\max(c)-\min(a)) \\
    c & \le & b + (1-r) \cdot (max(c)-\min(b))
\end{eqnarray*}
where $r \in [0,1]$ is a new auxiliary binary variable.

\subsubsection{The predicate {\tt int\_min}}
The semantics of {\tt int\_min(var int:~a, var int:~b, var int:~c)} is the equation 
$\min(a, b) = c$, i.e. the value of the integer variable~$c$ is the minimum
of the values of the integer variables~$a$ and~$b$. This can be represented by
the linear inequalities:
\begin{eqnarray*}
    c & \le & a \\
    c & \le & b \\
    a & \le & c + r \cdot (\max(a)-\min(c)) \\
    b & \le & c + (1-r) \cdot (max(b)-\min(c))
\end{eqnarray*}
where $r \in [0,1]$ is a new auxiliary binary variable.

\subsubsection{The predicate {\tt int\_mod}}
The semantics of {\tt int\_mod(var int:~n, var int:~d, var int:~r)} is 
$n \bmod d = r$, i.e. the value of the integer variable~$r$ is the modulo ---
i.e. the integer remainder of the integer division --- of the values of the integer variables~$n$ and~$d$. If $n$ is non-negative then $r$ is non-negative, too. 
The operator $\bmod$ returns the remainder (modulus) of the nominator~$n$ divided by 
the denominator~$d$. If $n$ is non-positive then $r$ is non-positive, too. 
This can be represented by
\begin{eqnarray*}
    {\tt int\_div(n, d, q)} \land d \cdot q + r = n
\end{eqnarray*}
(cf. Section~\ref{sec:fz:int_div}) where $q$ is a new auxiliary integer variable
and $r$ is the ``slack'' concerning the inequalities $d \cdot q + r \le n$ respective
$d \cdot q + r \ge n$ there.

\subsubsection{The Predicate {\tt int\_ne}}
The semantics of {\tt int\_ne(var int:~a, var int:~b)} is $a \ne b$, i.e. the 
value of the integer variable~$a$ is not equal to (different from) the value of 
the integer variable~$b$. This can be represented by linear inequalities:
\begin{eqnarray*}
    a & \le & b - 1 + r \cdot (\max(a) -\min(b) + 1) \\
    b & \le & a - 1 + (1-r) \cdot (\max(b) -\min(a) + 1) \\
\end{eqnarray*}
where $r \in [0,1]$ is a new auxiliary binary variable.

\subsubsection{The predicate {\tt int\_ne\_reif}}
The semantics of {\tt int\_ne\_reif(var int: a, var int: b, var bool: r)}
is $a \ne b \leftrightarrow r$, i.e. the value of the Boolean (binary) variable $r$ 
is 1 if the values of the integer variables~$a$ and~$b$ are not equal (different) and 0 
if they are equal. Following the approach in Section~\ref{sec:fz:int_lin_ne_reif} 
this can be represented by
\begin{eqnarray*}
    z & = & a - b \\
    z & = & r\cdot z \\
    r & \le & r \cdot |z|
\end{eqnarray*}
where $z \in [\min(a) - \max(b), \max(a) -\min(b)]$ is a new integer variable
and $|z| \in [l(|z|), u(|z|)]$ with
\begin{eqnarray*}
    l(|z|) & = & \left\{\begin{array}{rl}
                    0 & \mbox{if $\min(z) \le 0$} \\
                    \min(z) & \mbox{otherwise}
                \end{array}\right. \\
    u(|z|) & = &  \max(|\min(z)|, |\max(z)|) \enspace.
\end{eqnarray*}
For the representation of $|z|$ see Section~\ref{sec:fz:int_abs}.

\subsubsection{The predicate {\tt int\_plus}}
The semantics of {\tt int\_plus(var int:~a, var int:~b, var int:~c)} is
$a + b = c$ which is a linear equation.

\subsubsection{The predicate {\tt int\_pow}}
The semantics of {\tt int\_pow(var int:~x, var int:~y, var int:~z)} is
$z = x^y$ if $y \ge 0$ and $z = 1 \div x^{-y}$ if $y < 0$. In general this cannot be 
represented by polynomial systems and thus it is beyond the scope of our work.
However in the special case that $y$ is determined, i.e. $y = n$ respective $y \in \{n\}$
with $n>0$ then it holds $z =x^n$ which is polynomial. 
If $n$ is even, i.e. $n=2\cdot k$ we can set $z=u^2$ and $u=x^k$. If $n$ is odd, i.e.
$n=2\cdot k + 1$ we can set $z=x\cdot v$, $v=u^2$ and $u=x^k$. In both cases we can apply
this transformation steps recursively to $u=x^k$ until the considered result becomes
quadratic or linear.

\subsubsection{The predicate {\tt int\_times}}
The semantics of {\tt int\_times(var int:~a, var int:~b, var int:~c)} is 
$a \cdot b = c$ which is a binary variable product.

\subsection{Representing Bool(ean) FlatZinc Builtins}   

\subsubsection{The predicate {\tt array\_bool\_and}}\label{sec:fz:array_bool_and}
The semantics of the predicate {\tt array\_bool\_and(array [int] of var bool: as, var bool:~r)} is $r \leftrightarrow \bigwedge_{i} as_i$, i.e. the value of the binary 
variable~$r$ is 1 if all binary variables $as = as_1, \ldots, 
as_n$ are 1 and the value of~$r$ is 0 if at least one value of~$as_1, \ldots, as_n$ is 0. 
This can be represented by linear inequalities:
\begin{eqnarray*}
    r & \le & as_i \quad\mbox{for $i=1, \ldots, n$} \\
    r & \ge & \sum_{i=1}^n as_i - n + 1 \enspace.
\end{eqnarray*}

\subsubsection{The predicate {\tt array\_bool\_element}}
The semantics of the predicate {\tt array\_bool\_element(var\:int:\:i, array\:[int] of bool:~as, var bool:~c)} is $c = as_i$, i.e. the value of the binary variable $c$ 
is the $i$-th value in the binary values array~$as = as_1, \ldots, as_n$ where $i$ 
is an integer index variable $i$  with $i \in [\max(1,\min(i)), \min(n,\max(i))]$. 
In the special case that $as_{\min(i)} = \cdots = as_{\max(i)} = 1$ applies it must
hold that $c = 1$. In the special case that $as_{\min(i)} = \cdots = as_{\max(i)} = 0$ applies it must hold that $c = 0$. In all other cases this can be represented by linear equations, following Section~\ref{sec:fz:array_int_element}: 
\begin{eqnarray}
    1 & = & \sum_{j=\min(i)}^{\max(i)} b_j^i \label{eq:bool:one-hot_1} \\
    i & = & \sum_{j=\min(i)}^{\max(i)} j \cdot b_j^i  \label{eq:bool:one-hot_2} \\
    c & = & \sum_{j=\min(i)}^{\max(i)} as_j \cdot b_j^i \nonumber
\end{eqnarray}
where $b_{\min(i)}^i \in [0,1] , \ldots, b_{\max(i)}^i \in [0,1]$ are new auxiliary 
binary variables.

It should be noted that Equations~\ref{eq:bool:one-hot_1} and~\ref{eq:bool:one-hot_2}
define the \emph{one-hot-encoding} of the integer variable~$i$. This means that any 
other occurrence of~$i$ should be substituted/replaced by the sum
$\sum_{j=\min(i)}^{\max(i)} j \cdot b_j^i$ and any other binary encoding of~$i$ 
should be avoided.

\subsubsection{The predicate {\tt array\_bool\_xor}}
The semantics of {\tt array\_bool\_xor(array\:[int]\:of var\:bool:\:as)} 
is $\oplus_i as_i = 1$, i.e. exactly one value of the binary variables in
the array~$as = as_1, \ldots, as_n$ must be~1. This can be represented by a
linear equation:
\begin{eqnarray*}
    1 & = & \sum_{i=1}^n as_i \enspace.
\end{eqnarray*}

\subsubsection{The predicate {\tt array\_var\_bool\_element}}
The semantics of {\tt array\_var\_bool\_element(var int:~i, array [int] of var bool:~as,
var bool:~c)}  is $c = as_i$, i.e. the value of the binary variable $c$ is the $i$-th
value in the binary variable array~$as = as_1, \ldots, as_n$ where $i$ 
is an integer index variable $i$  with $i \in [\max(1,\min(i)), \min(n,\max(i))]$. 
Following Section~\ref{sec:fz:array_int_element} this can be represented by linear equations and binary variable products:
\begin{eqnarray}
    1 & = & \sum_{j=\min(i)}^{\max(i)} b_j^i \label{eq:var_bool:one-hot_1} \\
    i & = & \sum_{j=\min(i)}^{\max(i)} j \cdot b_j^i \label{eq:var_bool:one-hot_2} \\
    c & = & \sum_{j=\min(i)}^{\max(i)} z_j \nonumber \\
    z_j & = & as_j \cdot b_j^i \quad\mbox{for $j=\min(i), \ldots, \max(i)$} \nonumber
\end{eqnarray}
where $b_{\min(i)}^i \in [0,1] , \ldots, b_{\max(i)}^i \in [0,1]$ are new auxiliary 
binary variables and for $j=\min(i), \ldots, \max(i)$ the variables $z_{j} \in [\min(0, \min(a_{j})), \max(0,\max(a_{j}))]$  are new auxiliary integer variables.

It should be noted that Equations~\ref{eq:var_bool:one-hot_1} 
and~\ref{eq:var_bool:one-hot_2} define the \emph{one-hot-encoding} of the integer 
variable~$i$. This means that any other occurrence of~$i$ should be 
substituted/replaced by the sum $\sum_{j=\min(i)}^{\max(i)} j \cdot b_j^i$ 
and any other binary encoding of~$i$ should be avoided.

\subsubsection{The predicate {\tt bool2int}}
The semantics of {\tt bool2int(var bool:~a, var int:~b)} is $b \in [0,1]$ and
$a \leftrightarrow b = 1$, i.e. the value of integer variable~$b$ is~0 if the 
value of the Boolean variable~$a$ is \f{false} and the value of~$b$ is~1 if
the value of~$a$ is \f{true}. Due to the fact that we consider Boolean variables 
as binary integer variables, this constraint can be represented by a linear equation: 
\begin{eqnarray*}
    a & = & b
\end{eqnarray*}
(see considerations at the beginning of Section~\ref{sec:fz}).

\subsubsection{The predicate {\tt bool\_and}}
The semantics of {\tt bool\_and(var bool:~a, var bool:~b, var bool:~r)} is 
$r \leftrightarrow a \land b$. The value of the binary variable~$r$ is~0 if at
least one value of the binary variables~$a$ and~$b$ is zero and the value of~$r$ is~1 
if both values of~$a$ and~$b$ are 1. Following Section~\ref{sec:fz:array_bool_and}
this can be represented by linear inequalities: 
\begin{eqnarray*}
    r & \le & a \\
    r & \le & b \\
    a + b & \le & 1 + r
\end{eqnarray*}
or by the binary variable product:
\begin{eqnarray*}
    r & = & a \cdot b
\end{eqnarray*}
which requires that $r$ does not occur in the another variable product as multiplier.

\subsubsection{The predicate {\tt bool\_clause}}
The semantics of {\tt bool\_clause(array [int] of var bool:~as, array [int] of var 
bool:~bs)} is $1 = \bigvee_i as_i \lor \bigvee_j \lnot bs_j$ where $as = as_1, \ldots, 
as_n$ and $bs = bs_1, \ldots, bs_m$ are arrays of binary variables. This can be
represented by a linear inequality:
\begin{eqnarray*}
    1 & \le & \sum_{i=1}^n as_i + \sum_{j=1}^m (1-bs_j) \enspace.
\end{eqnarray*}

\subsubsection{The predicate {\tt bool\_eq}}
The semantics of {\tt bool\_eq(var bool:~a, var bool:~b)} is $a = b$, i.e. the value
of the binary variable~$a$ is equal to the value of the integer variable~$b$
which is a linear equation.

\subsubsection{The predicate {\tt bool\_eq\_reif}}
The semantics of {\tt bool\_eq\_reif(var bool:~a, var bool:~b, var bool:~r))}
is $a = b \leftrightarrow r$, i.e. the value of the Boolean (binary) variable $r$ 
is 1 if the values integer variables~$a$ and~$b$ are equal and 0 if they are
different. 
This can be represented by
\begin{eqnarray*}
   x & = & r \cdot a\\
   y & = & r \cdot b \\
   x & = & y \\
   (1-r) & = & (1-r)\cdot(a + b)
\end{eqnarray*}
where $x \in [0,1]$ and $y \in [0,1]$ are new auxiliary binary variables. 

\subsubsection{The predicate {\tt bool\_le}}
The semantics of {\tt bool\_le(var bool:~a, var bool:~b)} is $a \le b$, i.e. the value
of the binary variable~$a$ is less than or equal to the value of the binary variable~$b$
which is a linear inequality.

\subsubsection{The predicate {\tt bool\_le\_reif}}
The semantics of {\tt bool\_le\_reif(var bool:~a, var bool:~b, var bool:~r)}
is $a \le b \leftrightarrow r$, i.e. the value of the binary variable $r$ 
is~1 if the value of the binary variable~$a$ is less than or equal to the value
of the binary variable~$b$ and the value or~$r$ is~0 if the value of~$a$ is greater 
than the value of~$b$.
This can be represented by linear inequalities:
\begin{eqnarray*}
    a & \le & b + 1 - r \\
    b & \le & a - 1 + 2\cdot r \enspace.
\end{eqnarray*}

\subsubsection{The predicate {\tt bool\_lin\_eq}}
The semantics of {\tt bool\_lin\_eq(array\:[int]\:of~int~as, array\:[int]\:of var bool~bs,
var int:~c)} is $\sum_i as_i \cdot bs_i = c$, i.e. the sum of the values of 
the binary variables~$bs_1, \ldots, bs_n$ weighted by the integer values $as_1, \ldots, 
as_n$ --- assuming that both arrays have the same length~$n$ --- is equal to the 
value of the integer variable~$c$. This is a linear equation.

\subsubsection{The predicate {\tt bool\_lin\_le}}
The semantics of {\tt bool\_lin\_le(array\:[int]\:of int:~as, array\:[int]\:of var bool:~bs,
int:~c)} is $\sum_i as_i \cdot bs_i \le c$, i.e. the sum of the values of 
the binary variables~$bs_1, \ldots, bs_n$ weighted by the integer values $as_1, \ldots, 
as_n$ --- assuming that both arrays have the same length~$n$ --- is less than or equal to
the  integer value~$c$. This is a linear inequality.

\subsubsection{The predicate {\tt bool\_lt}}
The semantics of {\tt bool\_lt(var bool:~a, var bool:~b)} is $a < b$, i.e. the value
of the binary variable~$a$ is less than the value of the binary variable~$b$ which
means that $a=0$ and $b=1$ must hold, i.e. these are two simple linear equations.

\subsubsection{The predicate {\tt bool\_lt\_reif}}
The semantics of {\tt bool\_lt\_reif(var bool:~a, var bool:~b, var bool:~r)}
is $a < b \leftrightarrow r$, i.e. the value of the binary variable $r$ 
is~1 if the value of the binary variable~$a$ is less than the value of the binary 
variable~$b$ and the value or~$r$ is~0 if the value of~$a$ is greater than or equal 
to the value of~$b$.
This can be represented by linear inequalities:
\begin{eqnarray*}
    a & \le & b + 1 - 2\cdot r \\
    b & \le & a + r \enspace.
\end{eqnarray*}
or by a binary variable product:
\begin{eqnarray*}
    r & = & (1 - a) \cdot b
\end{eqnarray*}
which requires that $r$ does not occur in the another variable product as multiplier.

\subsubsection{The predicate {\tt bool\_not}}
The semantics of {\tt bool\_not(var bool:~a, var bool:~b)} is $a \ne b$, i.e. the 
value of the binary variable~$a$ is not equal to (different from) the value of 
the binary variable~$b$. This can be represented by a linear equation:
\begin{eqnarray*}
    a & = & 1-b \enspace.
\end{eqnarray*}

\subsubsection{The predicate {\tt bool\_or}}
The semantics of {\tt bool\_or(var bool: a, var bool: b, var bool: r)} is $a \lor b \leftrightarrow r$ i.e. the value of the binary variable $r$ is~1 if the value of at
least one of the binary variable~$a$ and~$b$ is 1 and the value or~$r$ is~0 if both
values of~$a$ and~$b$ are 0. This can be represented by linear inequalities:
\begin{eqnarray*}
    r & \le & a + b\\
    a + b & \le & 2\cdot r \enspace.
\end{eqnarray*}

\subsubsection{The predicate {\tt bool\_xor} (ternary)}
The semantics of {\tt bool\_xor(var bool:~a, var bool:~b, var bool:~r)} is $a \oplus b \leftrightarrow r$ i.e. the value of the binary variable $r$ is~1 if exactly one value 
of the binary variables~$a$ and~$b$ is 1 and the value or~$r$ is~0 if both
values of~$a$ and~$b$ are either~0 or~1.
This can be represented by
\begin{eqnarray*}
   (1-r)\cdot a & = & (1-r) \cdot b \\
   r \cdot (a + b) & = &  r
\end{eqnarray*}
and thus by linear equations, linear inequalities and binary variable products:
\begin{eqnarray*}
    x & = & r \cdot a \\
    y & = & r \cdot b \\
    a - x & = & b - y \\
    x + y & = & r
\end{eqnarray*}
where $x \in [0,1]$ and $y \in [0,1]$ are new auxiliary binary variables. 

\subsubsection{The predicate {\tt bool\_xor} (binary)}
The semantics of {\tt bool\_xor(var bool:~a, var bool:~b)} is $a \oplus b$ which
can be represented by a linear equation:
\begin{eqnarray*}
    a + b & = & 1 \enspace.
\end{eqnarray*}

\subsection{Representing Set FlatZinc Builtins}

Set constraints, more precisely finite integer set constraints are not completely
covered by our conversion from FlatZinc programs to Quadratic Integer Programs.
However, there are two set constraints over integer and binary variables and fixed 
sets of integer values which we will consider here.

\subsubsection{The predicate {\tt set\_in}}\label{sec:fz:set_in}
The semantics of {\tt set\_in(var int:~x, set of int:~S)} is $x \in S$.
This can be represented by linear equations:
\begin{eqnarray}
    1 & = & \sum_{s \in S \cap D(x)} b_S^x \label{eq:set:one-hot_1} \\
    x & = & \sum_{s \in S \cap D(x)} s \cdot b_s^x \label{eq:set:one-hot_2}
\end{eqnarray}
where $b_s^x \in [0,1]$ are new auxiliary binary variables.

It should be noted that Equations~\ref{eq:set:one-hot_1} and~\ref{eq:set:one-hot_2}
define the \emph{one-hot-encoding} of the integer variable~$x$. This should be 
respected when transforming the resulting QIP(FD) into a QUBO problem. 


\subsubsection{The predicate {\tt set\_in\_reif}}\label{sec:fz:set_in_reif}
The semantics of the predicate {\tt set\_in\_reif(var\:int:~x, set\:of:int:~S, 
var bool:~r)} is $r \leftrightarrow (x \in S)$, i.e. the value of the binary 
variable~$r$ is 1 if and only if the value of~$x$ is in the finite integer values 
set~$S$. This can be represented by linear equations:
\begin{eqnarray}
    1 & = & \sum_{j \in D(x)} b_j^x \label{eq:setreif:one-hot_1} \\
    x & = & \sum_{j \in D(x)} j \cdot b_j^x \label{eq:setreif:one-hot_2} \\
    r & = & \sum_{j \in D(x) \cap S} b_j^x \label{eq:setreif:one-hot_1} \nonumber
\end{eqnarray}
where $b_j^x \in [0,1]$ are new auxiliary binary variables.

It should be noted that Equations~\ref{eq:setreif:one-hot_1} 
and~\ref{eq:setreif:one-hot_2} define the \emph{one-hot-encoding} 
of the integer variable~$x$. This should be respected
when transforming the resulting QIP(FD) into a QUBO problem. 


\section{Conclusion and Future Work}

We have shown how Boolean and integer FlatZinc builtins can be represented
by linear equations, linear inequalities and/or binary variable products. This is the
basis for the transformation of finite domain integer FlatZinc programs via their
intermediate presenation as QIP(FD) into equivalent QUBO problems. 
This gives us the opportunity to transform a large class of MiniZinc programs into
QUBO problems and solve them also by Quantum Computing, e.g. with Quantum Annealers. 
Our future work will focus on the implementation of a MiniZinc-to-QUBO workflow.

\bibliographystyle{plain}
\bibliography{references}
\end{document}